\documentclass{article}

\usepackage{multicol,caption}
\usepackage{lipsum}

\usepackage{graphicx}
\usepackage{textcomp}
\usepackage{xcolor}
\usepackage[numbers]{natbib}
\usepackage[ruled,vlined]{algorithm2e}
\usepackage{algpseudocode}
\usepackage{bbm}
\usepackage{mathtools}
\usepackage{caption}
\usepackage{subcaption}
\usepackage{diagbox}
\usepackage{array}
\usepackage{amsmath}
\usepackage{amssymb}
\usepackage{amsfonts}
\usepackage{amsthm}
\usepackage{mathtools}

\usepackage{breqn}
\usepackage{enumitem}
\usepackage{physics}
\usepackage{thm-restate}
\usepackage{url}
\usepackage{tikz-cd}
\theoremstyle{definition}
\newtheorem{definition}{Definition}[section]
\theoremstyle{remark}

\DeclareMathOperator{\artanh}{artanh}
\newcommand{\comment}[1]{}
\allowdisplaybreaks
\usepackage{authblk}
\usepackage[margin=1.2cm]{geometry}
\usepackage{multicol}
\usepackage{blindtext}
\usepackage{epstopdf}
\begin{document}

\title{\textbf{Tight Differential Privacy Blanket for Shuffle Model}}

\author[1,2]{Sayan Biswas}
\author[1,2]{Kangsoo Jung}
\author[1,2]{Catuscia Palamidessi}

\affil[1]{{INRIA, France}}
\affil[2]{{LIX, \'{E}cole Polytechnique, France}}
\affil[ ]{{sayan.biswas@inria.fr, gangsoo.zeong@inria.fr, catuscia@lix.polytechnique.fr}}

\date{}

\setcounter{Maxaffil}{0}
\renewcommand\Affilfont{\itshape\small}
\maketitle

\begin{abstract}
With the recent bloom of focus on digital economy, the importance of personal data has seen a massive surge of late. Keeping pace with this trend, the model of data market is starting to emerge as a process to obtain high-quality personal information in exchange of incentives. To have a formal guarantee to protect the privacy of the sensitive data involved in digital economy, \emph{differential privacy (DP)} is the go-to technique, which has gained a lot of attention by the community recently. However, it is essential to optimize the privacy-utility trade-off by ensuring the highest level of privacy protection is ensured while preserving the utility of the data. In this paper, we theoretically derive sufficient and necessary conditions to have tight $(\epsilon,\,\delta)$-DP blankets for the shuffle model, which, to the best of our knowledge, have not been proven before, and, thus, characterize the best possible DP protection for shuffle models which can be implemented in data markets to ensure privacy-preserving trading of digital economy.
\end{abstract}

\begin{multicols}{2}

\section{Introduction and background}

As mentioned in \cite{wang2021managing}, privacy in digital economy is critical, especially for end-users who share their personal data. Differential privacy is a de-facto standard for privacy protection, however, it deteriorates the data utility. This trade-off between privacy and utility is a long standing problem in differential privacy.

An intermediate paradigm between the central and the local models of differential privacy (DP), known as the \emph{shuffle model}~\cite{bittau2017prochlo}, has recently gained popularity. As an initial step, the shuffle model uses a local mechanism to perturb the data individually like the local model of DP. After this local sanitization, a shuffler uniformly permutes the noisy data to dissolve their links with the corresponding data providers. This allows the shuffle model to achieve a certain level of DP guarantee using less noise than the local model ensuring that the shuffle model provides better utility than the local model whilst retaining the same advantages. Thus, the shuffle model has an advantage in the trade-off between privacy and utility for the digital economy.

The privacy guarantees provided by the  shuffle model have been rigorously studied by community of late and various results have been derived, both \emph{analytical} and \emph{numerical}. Obviously, analytical bounds have the advantage that they provide a concrete basis for reasoning and mathematically analysing properties such as privacy-utility trade-off. However, in the case of the shuffle model, most analytical bounds found in the literature are far from being tight. In this paper, we cover this gap and derive tight necessary and sufficient condition for having the tightest $(\epsilon,\,\delta)$-bounds for the DP guarantee provided by the shuffle model with the \emph{k-randomized response} ($k$-RR) local mechanism. 
We combine the idea of privacy blankets introduced by Balle et al. in \cite{balle2019privacy} and the concept of $(\epsilon,\,\delta)$-adaptive differential privacy (ADP) proposed by Sommer et al. in \cite{sommer2019privacy}.

\begin{definition}[Differential privacy~\cite{DworkDifferentialPrivacy,DworkApproximateDifferentialPrivacy}]
\label{def:dp}
For a certain query, a randomizing mechanism $\mathcal{K}$ taking datasets as input, provides \emph{($\epsilon,\,\delta)$-differential privacy (DP)} if for all neighbouring \footnote{\label{neighbouring dataset} differing in exactly one entry} datasets, $D_1$ and $D_2$, and all $S \subseteq$ Range($\mathcal{K}$), we have
\begin{equation*}
\mathbb{P}[\mathcal{K}(D_1) \in S] \leq e^{\epsilon}\,\mathbb{P}[\mathcal{K}(D_2) \in S] +\delta
\end{equation*}
\end{definition}

\begin{definition}[Adaptive differential privacy~\cite{sommer2019privacy}]
\label{def:adp}
Let us fix $x_0,\,x_1\,\in \mathcal{X}$, where $\mathcal{X}$ is the alphabet of the original (non-privatized) data, and let us fix a member $u$ in the dataset. For a certain query, a randomizing mechanism $\mathcal{K}$ provides \emph{$(\epsilon,\,\delta)$-adaptive differential privacy (ADP) for $x_0$ and $x_1$} if for all datasets, $D(x_0)$ and $D(x_1)$, and all $ S \subseteq$ Range($\mathcal{K}$), we have
\begin{equation*}
\mathbb{P}[\mathcal{K}(D(x_0)) \in S] \leq e^{\epsilon}\,\mathbb{P}[\mathcal{K}(D(x_1) \in S]\,+\,\delta
\end{equation*}
where $D(x_0)$ and $D(x_1)$ are datasets differing only in the entry of the fixed member $u$: $D(x)$ means that $u$ reports $x$ for every $x\,\in\,\mathcal{X}$, keeping the entries of all the other users the same.
\end{definition}

\begin{definition}[Tight DP (or ADP)~\cite{sommer2019privacy}]\label{def:tight}
    Let $\mathcal{K}$ be an $(\epsilon,\,\delta)$-DP (or ADP for chosen $x_0,\,x_1\in\mathcal{X}$) mechanism. We say that $\delta$ is \emph{tight} for $\mathcal{K}$ (wrt $\epsilon$ and $x_0,\,x_1$ in case of ADP) if there is no $\delta'<\delta$ such that $\mathcal{K}$ is $(\epsilon,\,\delta')$-DP (or ADP for $x_0,\,x_1$).
\end{definition}

\begin{definition}[Shuffle model~\cite{cheu2019distributed,erlingsson2019amplification}]
\label{def:shuffleseq}
Let $\mathcal{X}$ and $\mathcal{Y}$ be discrete alphabets for the original and the noisy data respectively. For any dataset of size $n\,\in\,\mathbb{N}$, the \emph{shuffle model} is defined as $\mathcal{M}:\mathcal{X}^n\mapsto\mathcal{Y}^n$, $\mathcal{M}=\mathcal{S} \circ \mathcal{R}^n$, where
\begin{itemize}
    \item[i)] $\mathcal{R}:\mathcal{X}\mapsto\mathcal{Y}$ is a local randomizer, stochastically mapping each element of the input dataset, sampled from $\mathcal{X}$, onto an element in $\mathcal{X}$, providing $\epsilon_0$-local differential privacy.
    \item[ii)] $\mathcal{S}:\mathcal{Y}^n\mapsto\mathcal{Y}^n$ is a shuffler that uniformly permutes the finite set of messages of size $n\,\in\,\mathbb{N}$, that it takes as an input.
\end{itemize} 

\end{definition}

\section{Tight differential privacy blanket analysis}

In \cite{balle2019privacy} Balle et al. introduced the concept of \emph{privacy blankets} for shuffle models, deriving some quantifiable privacy estimates for them in terms of $(\epsilon,\,\delta)$-DP. In this work, a notion of \emph{strong adversaries} was assumed, and the authors derived a precise condition for having $\epsilon\,\in\,\mathbb{R}^+$ and a corresponding space of $\delta\,\in\,\mathbb{R}^+$ that ensure a $(\epsilon,\,\delta)$-DP guarantee (Theorem 5.3 of \cite{balle2019privacy}. Although this work gives a straightforward and sufficient condition for having a space of $\epsilon$ and $\delta$ to encase the shuffle model with diffenrential privacy guarantees, it is crucial to note that no explicit theoretical condition exists that can ensure the tightness\footnote{as in Definition \ref{def:tight}} of such privacy blankets. 

In the same year, Sommer et al. in \cite{sommer2019privacy} gave an explicit condition for having a $\delta$ for a chosen $\epsilon$ that would foster a tight $(\epsilon,\,\delta)$-ADP, for a couple of given inputs, on any probabilistic mechanism (Lemma 5 in \cite{sommer2019privacy}).

For comparing utilities between the shuffle model and other privacy mechanisms under a certain level of privacy, having a slack $(\epsilon,\,\delta)$-privacy coverage would not suffice as we would not be able to guage the best privacy guarantee provided without having the tight bounds. Therefore, in this paper, we have investigated and analysed the explicit conditions under which the result of Balle et al. in \cite{balle2019privacy}) improves and ensures a tight privacy blanket for the shuffle model using $k$-RR mechanism as its local randomizer. In particular, drawing parallels between Lemma 5 of \cite{sommer2019privacy}, we derived a necessary and sufficient condition for Theorem 5.3 of \cite{balle2019privacy} to ensure a tight $(\epsilon,\,\delta)$-DP guarantee.
 
We analyze the two cases of Lemma 5 in \cite{sommer2019privacy} separately:

\begin{align}\scriptsize
    \text{\emph{Case 1}: }\scriptsize\frac{1}{e^{\epsilon_0}} \leq \frac{(e^\epsilon-1)^2}{(e^\epsilon+1)^2(e^{\epsilon_0}-e^{-\epsilon_0})^2} \label{EasyCase}\\
    \scriptsize\text{\emph{Case 2}: }\frac{1}{e^{\epsilon_0}} > \frac{(e^\epsilon-1)^2}{(e^\epsilon+1)^2(e^{\epsilon_0}-e^{-\epsilon_0})^2}\label{DifficultCase}
\end{align}
\par

Setting $C=1-e^{-2} \approx 0.86$, Lemma 5 in \cite{sommer2019privacy}) proves that $\mathcal{M}$ guarantees $(\epsilon,\,\delta)$-DP for every choice of $\epsilon\in\mathbb{R}^+$ and $\delta\in\mathbb{R}^+$ satisfying:

\begin{align}\label{eq:balle}\scriptsize
    \delta\geq 
    \begin{cases}
        \frac{(e^{\epsilon}+1)^2(e^{\epsilon_0}-e^{-\epsilon_0})^2}{4n(e^\epsilon-1)}e^{-Cn\frac{1}{e^{\epsilon_0}}} & \text{ for \emph{Case 1}.}\\
        \frac{(e^{\epsilon}+1)^2(e^{\epsilon_0}-e^{-\epsilon_0})^2}{4n(e^\epsilon-1)}e^{-Cn\frac{(e^\epsilon-1)^2}{(e^\epsilon+1)^2(e^{\epsilon_0}-e^{-\epsilon_0})^2}} & \text{ for \emph{Case 2}.}
    \end{cases}
\end{align}
\par

Note that for a given $\epsilon>0$, if $\delta_1$ and $\delta_2$ both satisfy Lemma 5~\cite{sommer2019privacy}, then $\max\{\delta_1,\delta_2\}$ will provide a slack privacy blanket, making it hard to quantify a definite differential privacy level for $\mathcal{M}$. It is imperative to have a tight privacy parameter in order to proceed with any sort of privacy-utility analysis and lay down a notion of comparison between $\mathcal{M}$ and other DP mechanisms. Hence, to investigate the existence of a tight $\delta$ for a chosen $\epsilon$ and to examine the precise conditions under which Result 1 by Balle et al. can provide a tight differential privacy guarantee for $\mathcal{M}$, we consider the minimum $\delta$ satisfying \eqref{eq:balle} for a given $\epsilon$ and equate it to Lemma 5 of \cite{sommer2019privacy}) and maximize it over all pairs of input values to translate the ADP guarantee of Sommer et al.'s result to the standard notion of DP.

As we are interested to examine if we can find $\epsilon>0$ and, correspondingly, $\delta>0$ that provide a tight DP guarantee for $\mathcal{M}$, we define the constants $\kappa_i,i \in \{1,\ldots,8\}$, that will become handy to simplify the mathematical results derived and used in the paper, as follows:

\small
\begin{align}
   \kappa_1\coloneqq\frac{(e^{\epsilon_0}-e^{-\epsilon_0})^2e^{Cne^{-\epsilon_0}}}{4} \label{kappa1}\\
   \kappa_2\coloneqq1+\frac{e^{\epsilon_0}-1}{n+(e^{\epsilon_0}-1)\pi(x_0)(n-1)}\label{kappa2}\\
   \kappa_3\coloneqq\frac{n+(e^{\epsilon_0}-1)(n\pi(x_0)+1-\pi(x_0))}{e^{\epsilon_0}+k-1}\label{kappa3}\\
   \kappa_4\coloneqq2\artanh{\left(\frac{2\sinh{(\epsilon_0)}}{e^{\epsilon_0/2}}\right)}\label{kappa4}\\
   \kappa_5\coloneqq\frac{\sinh^2{\epsilon_0}}{n}\label{kappa5}
\end{align}
\normalsize

\begin{definition}[Critical Polynomial]
For a given privacy parameter, $\epsilon_0$, of the $k$-RR mechanism used in $\mathcal{M}$ and $n$ samples drawn from $\mathcal{X}$ following a distribution $\pi$, let $P_1(x),P_2(x)\in\mathbb{R}[x]$ be polynomials defined as:
\small
\begin{equation*}
    P_1(x)\coloneqq\kappa_1\kappa_2(x+1)^2, \,P_2(x)\coloneqq\kappa_3(\kappa_2-x)(x-1)   
\end{equation*}
\normalsize
Let $P(x)\coloneqq P_1(x)-P_2(x)$. We call $P(x)$ to be our \emph{critical polynomial}.
\end{definition}

\begin{definition}[Critical Equation]
Let the \emph{critical equation} $H(x)$ be defined as:
\begin{equation*}
    H(x)\coloneqq2\kappa_5\kappa_2e^{-\frac{Cx^2}{4\kappa_5}}+x^2\kappa_3(\kappa_2+1)-x\kappa_3(\kappa_2-1)
\end{equation*}
\end{definition}

\begin{restatable}{theorem}{asymptotictightprivacyone}\label{th:case1}
For $\epsilon>\max\{\kappa_4,\ln\kappa_2\}$, taking $\delta$ as in \eqref{eq:balle} asymptotically provides tight $(\epsilon,\delta)$-ADP guarantee for $\mathcal{M}$ wrt $x_0,\,x_1$ as $n\to\infty$.
\end{restatable}

\begin{restatable}{theorem}{asymptotictightprivacytwo}\label{th:case2}
For $\ln{\kappa_2}<\epsilon<\kappa_4$, taking $\delta$ as in \eqref{eq:balle} asymptotically provides a tight $(\epsilon,\,\delta)$- ADP blanket for $\mathcal{M}$ wrt $x_0,\,x_1$, as $n\to\infty$.
\end{restatable}

For any $x_i\in\mathcal{X}$ as the primary input of $u$, let $\mathcal{S}_1(x_i)$ denote the space $[\kappa_4,\ln{\kappa_2(x_i)})\subset\mathbb{R}$ and $\mathcal{S}_2(x_i)$ denote the space $(0,\min\{\kappa_4,\ln{\kappa_2(x_i)}\})\subset\mathbb{R}$.

\begin{restatable}{theorem}{commonspace}\label{DPguarantee}
\begin{enumerate}[label=\alph*)]
    \item If $\kappa_4<\ln{\kappa_2(x_i)}$ for all $x_i\in\mathcal{X}$, and if $P(x)=0$ has a real solution in $[e^{\kappa_4},\kappa_2)$, then $\mathcal{S}_1\coloneqq\bigcap\limits_{x_i\in\mathcal{X}}\mathcal{S}_1(x_i)\neq\phi$. Moreover, for every $\epsilon\in\mathcal{S}_1$, choosing $\delta$ as in \eqref{eq:balle} ensures a tight $(\epsilon,\,\delta)$-DP blanket for $\mathcal{M}$.
    
    \item If $H(x)=0$ has a real solution in $(0,\mu)$, where $\mu=\min\left\{\tanh{(\frac{\kappa_4}{2})},\tanh{(\frac{\ln{\kappa_2(x_i)}}{2})}\right\}$, then setting $\mathcal{S}_2\coloneqq\bigcap\limits_{x_i\in\mathcal{X}}\mathcal{S}_2(x_i)\neq\phi$. Moreover, for every $\epsilon\in\mathcal{S}_2$, choosing $\delta$ ensures a tight $(\epsilon,\,\delta)$-DP blanket for $\mathcal{M}$. 
\end{enumerate}
\end{restatable}

\section{Discussion and future work}
The theoretical conditions derived in this work to make Balle et al.'s bounds for privacy blankets of shuffle models give us an analytical insight to the cases when we could obtain the best privacy guarantee for shuffle models using $k$-RR local randomizers. Studying the space of the tight DP guarantees for the shuffle models could be a possible breakthrough in this area, as this could then be used to tune the hyperparamters such as the privacy level of the local randomizer ($\epsilon_0$) and the number of samples ($n$) for implementing the shuffle model in various areas of privacy-preserving data analysis. Naturally, this would also help to maximize the utility for a given level of privacy, once the best DP guarantee for the shuffle model is illustrated, aiding to resolve the privacy-utility trade-off for shuffle models of DP.

We plan on studying more generalized forms of shuffle models using different local randomizers and comparing their utilities with the central models. Also, wish to extend this work in the context of digital markets by implementing shuffle models and exploiting its DP guarantees in the context of differentially private data market and privacy pricing~\cite{jung2021establishing, biswas2021incentive}. In \cite{jung2021establishing}, Jung et al. proposed a federated data trading framework in which data providers coalesce to form a federations to increase their bargaining power in data trading. An immediate extension would be to analyse the mechanism of \cite{jung2021establishing} under the environment of shuffle models implement by each federation and study a data pricing mechanism based on the privacy amplification by shuffling.

\bibliographystyle{IEEEtran}
\bibliography{reference}

\end{multicols}

\end{document}